\documentclass[article]{elsarticle}
\usepackage{lineno,hyperref}
\modulolinenumbers[5]

\usepackage{amsmath}
\usepackage{amssymb}

\journal{Annals of Physics}

\newcommand{\apjl}{Astrophys. J. Lett.}%
\newcommand{\aap}{Astron. Astrophys.}%
\newcommand{\mnras}{Mon. Not. Roy. Astron. Soc.}%
\newcommand{\lrr}{Living Reviews in Relativity}%
\newcommand{\prl}{Phys. Rev. Lett.}%
\newcommand{\apj}{Astrophys. J.}%
\newcommand{\prd}{Phys. Rev. D}%
\newcommand{\nat}{Nature}%
\newcommand{\physrep}{Physics Reports}%

\begin{document}

\begin{frontmatter}

\title{Equation of state constraints from multi-messenger observations of neutron star mergers}

\author{Andreas Bauswein}
\address{GSI Helmholtzzentrum f\"ur Schwerionenforschung, Planckstra{\ss}e 1, 64291 Darmstadt, Germany}
\address{Heidelberg Institute for Theoretical Studies, Schloss-Wolfsbrunnenweg 35, 69118 Heidelberg, Germany}

%
\ead{a.bauswein@gsi.de}
%

\begin{abstract}
The very first detection of gravitational waves from a neutron star binary merger, GW170817, exceeded all expectations. The event was relatively nearby, which may point to a relatively high merger rate. It was possible to extract finite-size effects from the gravitational-wave signal, which constrains the nuclear equation of state. Also, an electromagnetic counterpart was detected at many wavebands from radio to gamma rays marking the begin of a new multi-messenger era involving gravitational waves. We describe how multi-messenger observations of GW170817 are employed to constrain the nuclear equation of state. Combining the information from the optical emission and the mass measurement through gravitational waves leads to a lower limit on neutron star radii. According to this conservative analysis, which employs a minimum set of assumptions, the radii of neutron stars with typical masses should be larger than about 10.7~km. This implies a lower limit on the tidal deformability of about 210, while much stronger lower bounds are not supported by the data of GW170817. The multi-messenger interpretation of GW170817 rules out very soft nuclear matter and complements the upper bounds on NS radii which are derived from the measurement of finite-size effects during the pre-merger phase. We highlight the future potential of multi-messenger observations and of GW measurements of the postmerger phase for constraining the nuclear equation of state. Finally, we propose an observing strategy to maximize the scientific yield of future multi-messenger observations.
\end{abstract}


\end{frontmatter}


\section{Introduction}
The fact that some number of neutron star (NS) binaries merge, had been widely anticipated in the scientific community before GW170817. This expectation was mostly based on observations of double NS systems containing at least one pulsar. Precise pulsar timing allows the determination of various orbital parameters, which led to the conclusion that a fraction of NS binaries would collide as a consequence of gravitational-wave (GW) emission within less than a Hubble time. Only a minority of scientists expected the observation of a NS merger in the year 2017 even after several detections of black-hole binaries succeeded in the previous years\footnote{This was at least the perception of the author of this article.}. Many researchers considered a NS merger much closer than 100~Mpc to be very unlikely. However, Nature was very kind and placed a NS merger at about 40~Mpc from Earth right during the second LIGO/Virgo observing run~\cite{Abbott2017}. The analysis of the GW signal revealed the binary parameters of the system. Specifically, the so-called chirp mass~$\mathcal{M}=(M_1 M_2)^{3/5}(M_1+M_2)^{-1/5}$ was very accurately measured, while the binary mass ratio $q=M_1/M_2$ was only roughly determined to be in the range between 0.7 and 1. Combining this information implies that GW170817 was very likely a merger of two NSs unless some exotic mechanism could produce stellar black holes with masses smaller than the maximum mass of NSs. The total mass of GW170817 was found to be $M_\mathrm{tot}=2.74^{+0.04}_{-0.01}~M_\odot$ consistent with typical masses observed in NS binaries~\cite{Lattimer2012}. 

The localization of the event through the GW signal also allowed to find an electromagnetic counterpart in the optical/infrared about half a day after the merger (electromagnetic radiation was observed throughout the full electromagnetic spectrum from radio to gamma rays~\cite{Abbott2017b}). The electromagnetic emission at optical/infrared wavelengths evolved on the time scale of days and was observable for more than a week, e.g.~\cite{Abbott2017b,Villar2017}. The importance of these measurements lies in the fact that they provide the to date best evidence that the ejecta of NS mergers undergo the rapid neutron capture process (r-process) to form heavy elements. This conclusion is based on the interpretation of the light curve, which is compatible with matter being heated by the radioactive decays in the aftermath of the r-process. Although estimates of ejecta properties like their mass and outflow velocities depend on the underlying modeling (see~\cite{Cote2018} for a compilation of different results), the follow-up observations of GW170817 combined with approximate rate estimates indicate that NS mergers play an important role for the enrichment of the Universe by r-process elements. This is a remarkable development considering that only in the last $\sim 10$ years NS mergers have received increasing attention as potential sources of heavy r-process elements~\cite{Metzger2010,Roberts2011,Goriely2011} with some earlier works starting in the 70ies~\cite{Lattimer1974,Eichler1989,Ruffert1997,Freiburghaus1999,Rosswog1999}. 

Another important development after GW170817 are observational constraints on the incompletely known equation of state (EoS) of high-density matter, which uniquely determines stellar parameters of NSs likes their radii and their tidal deformability, i.e. the response to an external tidal field. Two types of constraints have been presented. 1) Considering solely the GW signal, the measurement of finite-size effects during the late inspiral phase constrains the tidal deformability from above. While the precise limits depend somewhat on assumptions made for the analysis, the GW signal establishes a robust upper bound on NS radii of about 13.5~km for typical NS masses and thus excludes very stiff nuclear matter. 2) A second type of constraints makes use of the additional information which is provided by the electromagnetic signature of GW170817. Based on the multi-messenger observations of GW170817 a variety of such studies has been put forward~\cite{Margalit2017,Bauswein2017,Shibata2017,Rezzolla2018,Radice2018,Ruiz2018,Coughlin2018,Radice2019,Koeppel2019,Kiuchi2019}. 
Some of these EoS constraints are more tentative, while others are more robust and include a proper error analysis like the one described below. Interestingly, the multi-messenger interpretation of GW170817 discussed in Sect.~\ref{sec:main} provides a lower bound on NS radii and is thus complementary to the limits given by finite-size effects during the inspiral. Finally, both types of constraints have lead to a number of studies discussing which specific EoS models are compatible with the new observations, e.g.~\cite{Fattoyev2018}. We emphasize that because of the limited space we cannot mention all constraints and studies in the context of GW170817 and we can only provide an incomplete list of the relevant literature.

All these different considerations may only be a first indication of the potential of future multi-messenger observations which can be expected in the next years.

\section{EoS constraints from GW170817}\label{sec:main}

Here we describe how the multi-messenger observations of GW170817 provide a lower limit on NS radii and the nuclear EoS with a minimum of assumptions. More details of this idea can be found in~\cite{Bauswein2017}. The constraint is based on three main arguments: (a) the assumption that GW170817 did not result in a direct gravitational collapse of the merger remnant, (b) an empirical relation for the threshold binary mass for direct black-hole formation, which depends on NS radii and the maximum mass of nonrotating NSs, and (c) causality, which implies certain constraints on NS properties. We discuss some details before combining the arguments to derive NS radius constraints.

\subsection{No direct collapse in GW170817}
Different groups observed the electromagnetic counterpart of GW170817 in the optical and infrared, which evolved on the time scale of days (e.g.~\cite{Villar2017}). The light curve is compatible with ejecta from a NS merger, which is heated by radioactive decays during the r-process. Fitting the observed luminosity to theoretical emission models allows to infer the ejecta masses and other properties of the outflow~\cite{Metzger2017a}. There is general agreement between the different models that the emission is best explained by the ejection of a few hundreds of a solar mass of matter undergoing the r-process. There is no detailed information about the composition of the ejecta available (but there may be evidence of spectral features~\cite{Smartt2017}). The emission points to different components of the ejecta, which is consistent with theoretical expectations that there are dynamical ejecta getting unbound within the first milliseconds after merging followed by a slower component which emerges on somewhat longer time scales by viscous, nuclear and neutrino processes. Although details remain model-dependent, the overall agreement with theoretical predictions provides strong evidence that the r-process took place in the outflow of GW170817 and that a few $0.01~M_\odot$ became gravitationally unbound in this event.
\begin{figure}
\includegraphics[width=0.8\columnwidth]{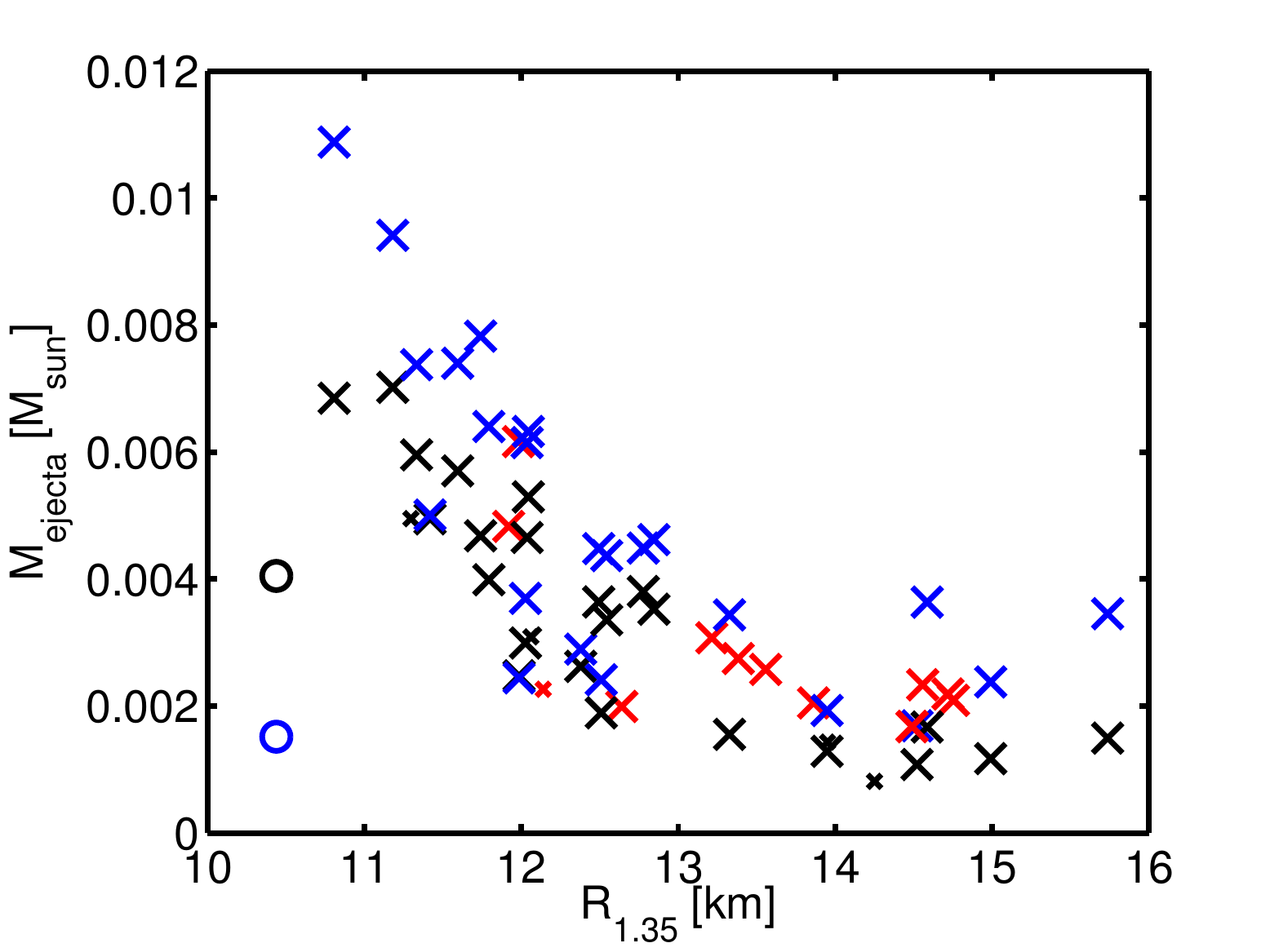}
\caption{Dynamical ejecta mass as function of the radius $R_{1.35}$ of a 1.35~$M_\odot$ NS for different EoS models. All data points refer to merger simulations with two equal-mass NSs of 1.35~$M_\odot$, thus a system with a total mass comparable to that of GW170817. The open circles indicate simulations which lead to a prompt collapse of the merger remnant. Red symbols stand for fully temperature-dependent EoS models, wheres balc and blue symbols mark data from simulations with an approximate treatment of thermal effects. See~\cite{Bauswein2013a} for details. Figure taken from~\cite{Bauswein2013a}. See~\cite{Bauswein2013a} for a similar plot for 1.2-1.5~$M_\odot$ mergers, which generally yield higher ejecta masses.}
\label{fig:mejr135}
\end{figure}

\begin{figure}
\includegraphics[width=0.8\columnwidth]{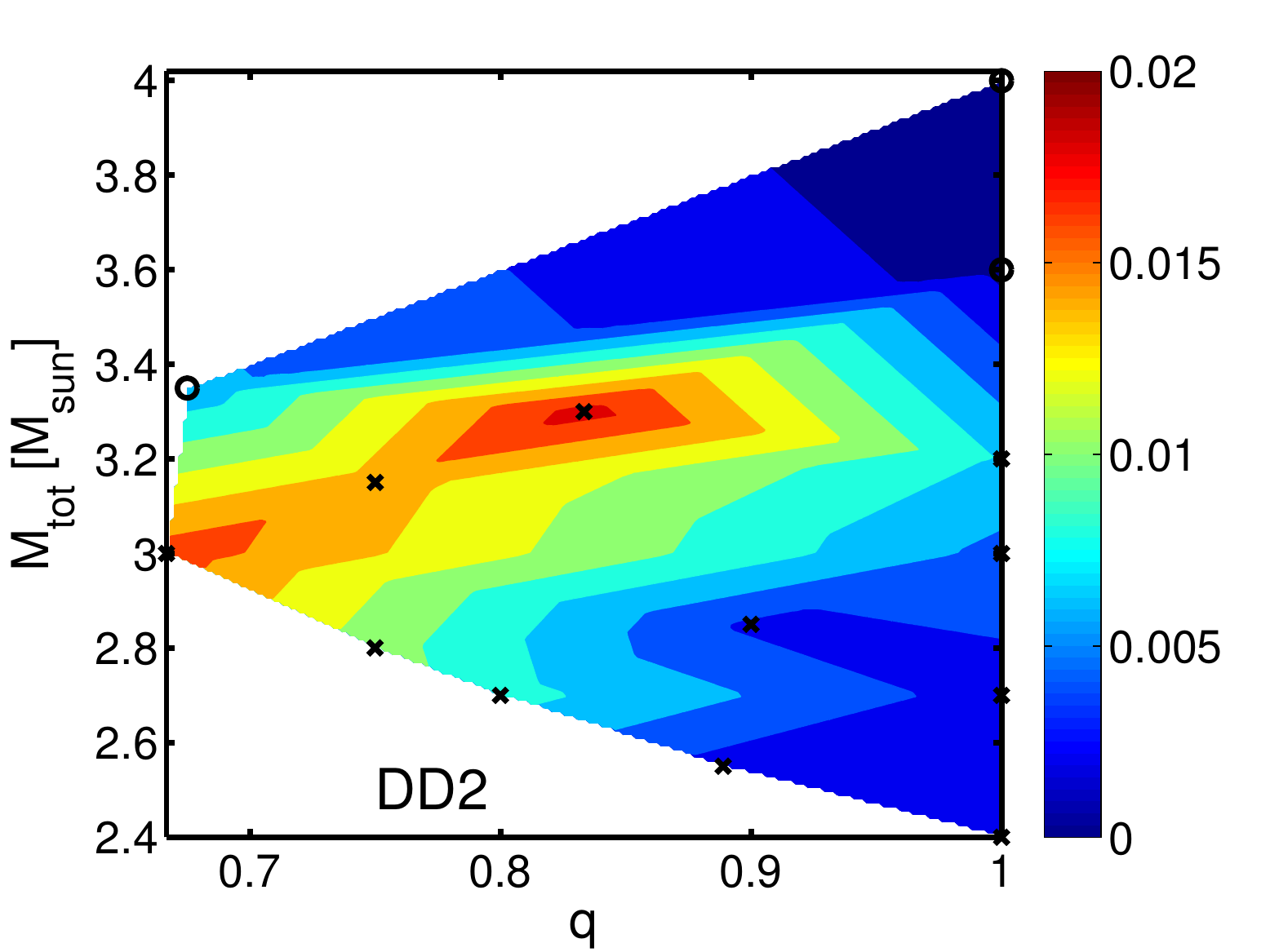}
\caption{Dynamical ejecta mass (color-coded in solar masses) as function of the binary mass ratio $q=M_1/M_2$ and the total binary mass $M_\mathrm{tot}=M_1+M_2$ for the DD2 model EoS~\cite{Hempel2012}. The open circles indicate simulations which lead to a prompt collapse of the merger remnant. Crosses mark simulations resulting in a NS remnant. Figure taken from~\cite{Bauswein2013a}.}
\label{fig:mtotq}
\end{figure}

This amount of ejecta is at the high end of what is expected from hydrodynamical simulations~\cite{Hotokezaka2013,Bauswein2013a,Metzger2014,Just2015}, keeping in mind that one cannot easily distinguish dynamical and secular ejecta which can be comparable in mass (see e.g. the compilation by~\cite{Wu2016}). One should also appreciate that determining the dynamical or secular ejecta mass in simulations comes with an uncertainty of at least a few 10 per cent. From simulations it is known that the ejecta mass depends on the binary masses and the incompletely known high-density EoS~\cite{Hotokezaka2013,Bauswein2013a} (see Figs.~\ref{fig:mejr135} and~\ref{fig:mtotq}). Figure~\ref{fig:mejr135} shows the dynamical ejecta mass for 1.35-1.35~$M_\odot$ mergers for many different model EoSs. The different models are characterized by the radius of a 1.35~$M_\odot$ NS described the different EoSs, where soft EoSs yield small radii and stiff nuclear matter results in larger radii. It is interesting to note that the amount of dynamical ejecta critically depends on the immediate outcome of the merger. If the merging leads to a prompt gravitational collapse on a dynamical time scale, the ejecta mass is significantly reduced compared to the case where a NS merger remnant forms which survives for at least a few milliseconds or longer (see circles in Fig.~\ref{fig:mejr135} and Fig.~\ref{fig:mtotq}). Also asymmetric binaries with a mass ratio different from unity lead to relatively small ejecta masses if the remnant undergoes a prompt collapse. Fig.~\ref{fig:mtotq} shows in more detail how binary parameters affect the amount of dynamical ejecta. Mass ejection is generally enhanced for asymmetric systems because tidal ejection of matter is stronger. The dynamical ejecta is reduced for low-mass systems likely because the merger proceeds in a less violent way compared to binaries with higher total binary mass which do not undergo a prompt collapse.

For a given EoS the type of merger remnant (black hole or NS) critically depends on the total binary mass. If the total binary $M_\mathrm{tot}$ mass exceeds a certain threshold mass $M_\mathrm{thres}$, the remnant cannot be stabilized against the gravitational collapse and directly forms a black hole. The mass ratio of the binary for a fixed total mass has only a very weak impact, which has been shown in for limited set of simulations~\cite{Bauswein2013,Bauswein2017}. Obviously, the threshold binary mass $M_\mathrm{thres}$ for prompt black-hole formation depends sensitively on the EoS because the exact properties of high-density matter like its stiffness determine how much mass can be supported against the gravitational attraction. We discuss this point in more detail below. The reduction of the dynamical ejecta by a prompt collapse is generically found for all tested EoSs. 

In essence, the relatively high ejecta mass inferred from GW170817 indicates that there was no direct black-hole formation in this event. We will use this conclusion as working hypothesis to derive NS radius constraints. Although there is no final proof for this assumption, it appears reasonable, as also other arguments point towards no prompt black-hole formation. For instance, Ref.~\cite{Ruiz2017} showed in general-relativistic magneto-hydrodynamical simulations that jet formation only takes place if the NS remnant undergoes a delayed gravitational collapse, while prompt black-hole formation may not lead to a beamed relativistic outflow. Given the observational indications for a relativistic jet in GW170817, this may provide another piece of evidence for no direct collapse of the merger remnant in GW170817.

The main conclusion from the discussion above is that it is reasonable to assume that there was no direct collapse in GW170817. Therefore, the measured total binary mass $M_\mathrm{tot}^\mathrm{measured}=2.74^{+0.04}_{-0.01}~M_\odot$ is smaller than the threshold binary mass $M_\mathrm{thres}$ for prompt black-hole formation.

\subsection{Empirical relation for $M_\mathrm{thres}$}
By means of a large set of hydrodynamical merger simulations (see~\cite{Bauswein2013}) it is possible to infer an approximate empirical relation describing the EoS dependence of the threshold binary mass for prompt black-hole formation. To this end, we performed simulations for different total binary masses and determined the outcome. Like this one can identify $M_\mathrm{thres}$ for every model EoS and relate  $M_\mathrm{thres}$ to certain properties of the EoS, e.g. stellar parameters of nonrotating NSs. It is found that generally  $M_\mathrm{thres}$ increases with the maximum mass $M_\mathrm{max}$ of nonrotating NSs and with radii of nonrotating NSs. For instance, the threshold binary mass can be described to good accuracy by
\begin{equation}\label{eq:mthrrmax}
M_\mathrm{thres}=\left(-3.38\frac{G\,M_\mathrm{max}}{c^2\,R_\mathrm{max}} +2.43 \right)\,M_\mathrm{max}
\end{equation}
or by
\begin{equation}\label{eq:mthrr16}
M_\mathrm{thres}=\left(-3.606\frac{G\,M_\mathrm{max}}{c^2\,R_{1.6}} +2.38 \right)\,M_\mathrm{max}
\end{equation}
with $R_\mathrm{max}$ being the radius of the maximum-mass configuration and $R_{1.6}$ being the radius of a 1.6~$M_\odot$ NS ($G$ and $c$ are the gravitational constant and the speed of light, respectively). These equations represent empirical relations based on hydrodynamical models for a representative set of EoSs, which should be accurate to within a few per cent with uncertainties resulting from the finite spacing of merger configurations in the space of total binary masses and from the underlying physical and numerical model. Note the generally good agreement with the recent results from~\cite{Koeppel2019} for a smaller set of EoSs using grid-based simulations in full General Relativity. Note that these general dependencies have been corroborated by semi-analytical models based on stellar equilibrium configurations~\cite{Bauswein2017a}.

\begin{figure}
\includegraphics[width=0.8\columnwidth]{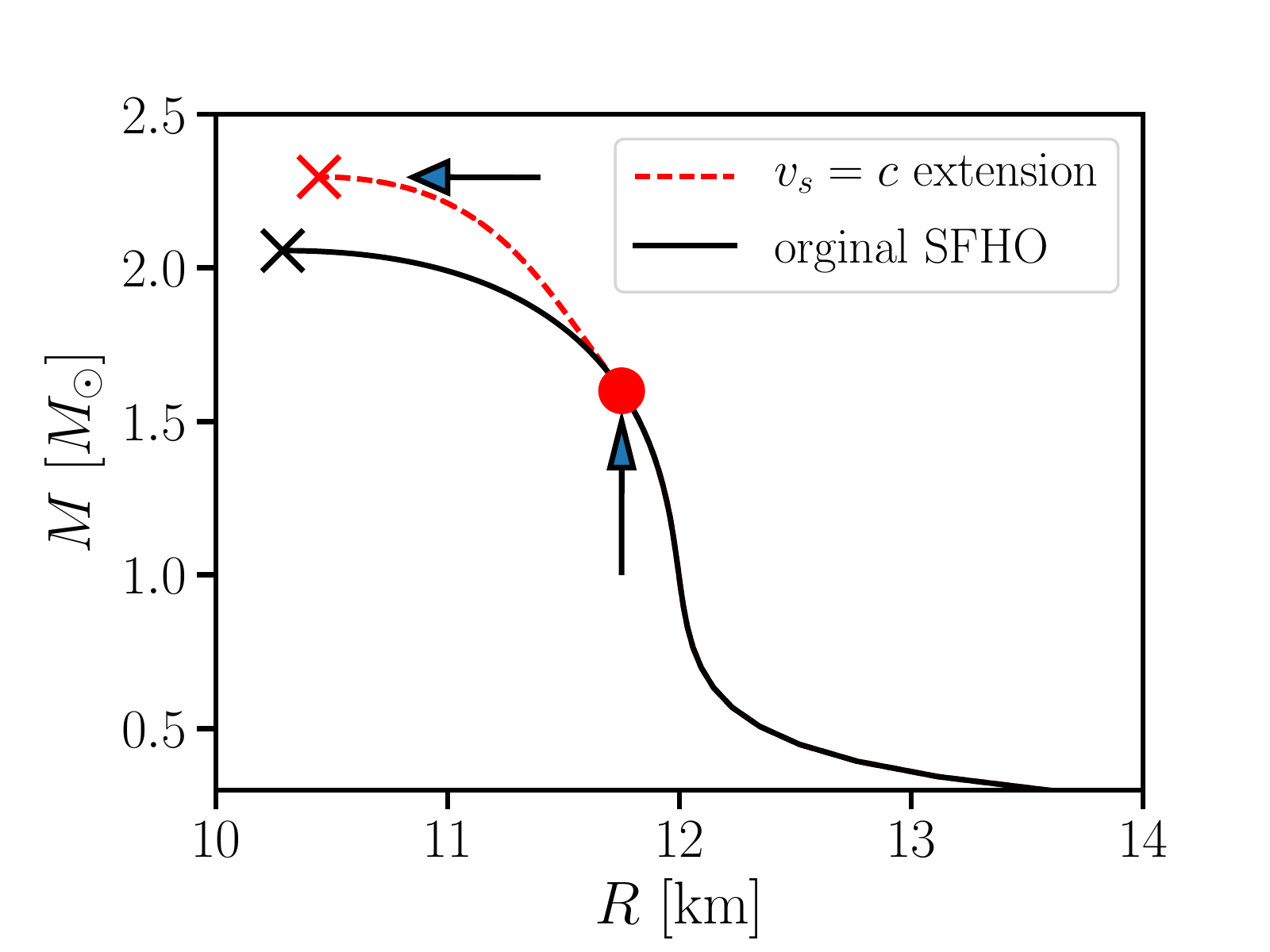}
\caption{Example for a mass-radius relation of an artificially modified EoS at higher densities. The original model is shown by the black curve. This model EoS is modified by assuming that the EoS becomes maximally stiff beyond the central density of a 1.6~$M_\odot$, i.e. that the speed of sound equals the speed of light. This artificial modification alters the mass-radius relation beyond a mass of 1.6~$M_\odot$ and leads to the red dashed curve. The resulting maximum mass (red cross) is approximately the highest possible maximum mass that an EoS with the given radius $R_{1.6}$ at 1.6~$M_\odot$ (red dot) could have.}
\label{fig:causalsfho}
\end{figure}

\begin{figure}
\includegraphics[width=0.8\columnwidth]{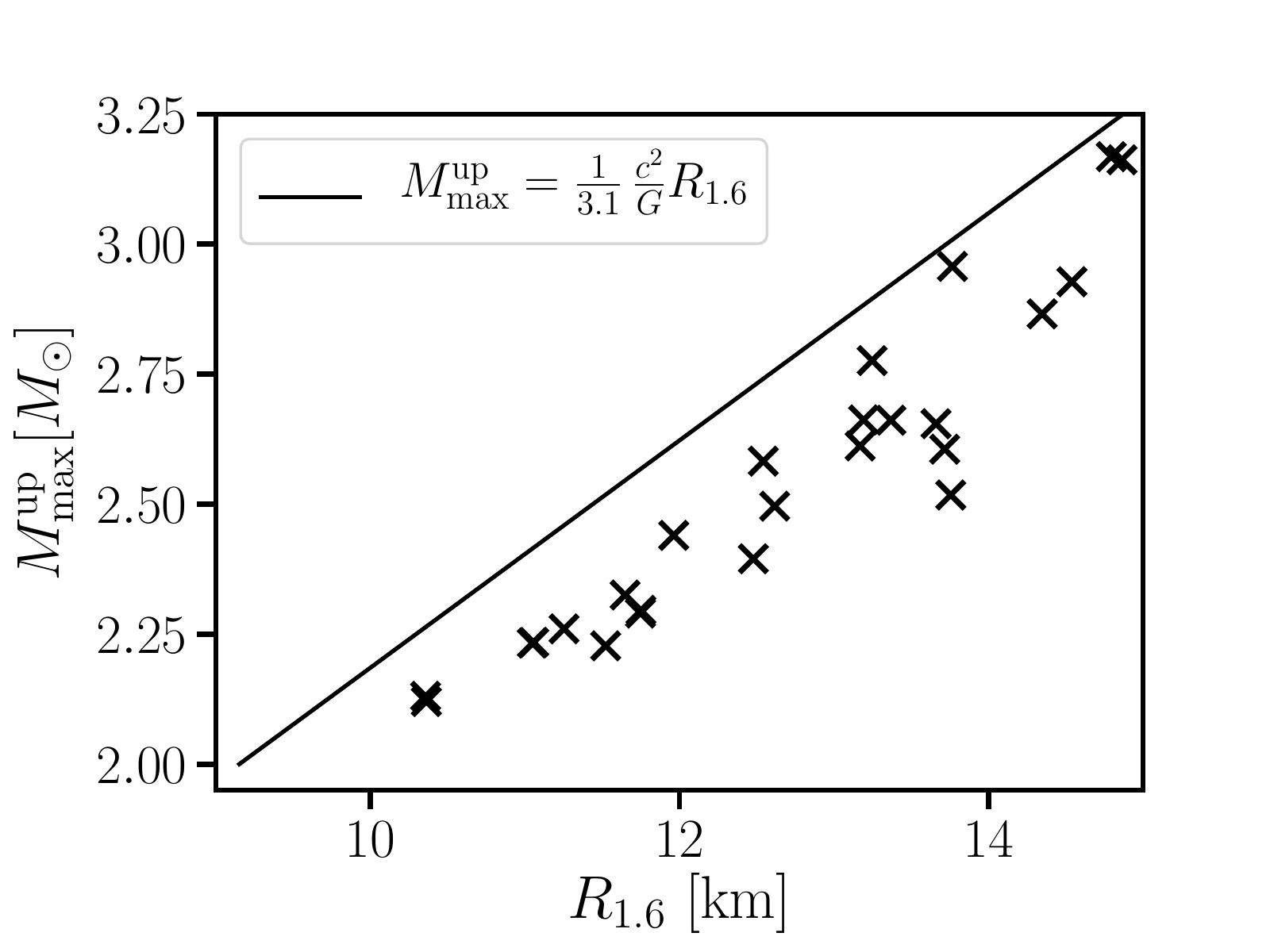}
\caption{Upper limit $M_\mathrm{max}^\mathrm{up}$ on the maximum mass of nonrotating NSs as function of the radius $R_{1.6}$ of a NS with 1.6~$M_\odot$. Data points are generated by modifying the high-density part of a large sample of microphysical EoS models as illustrated by the example in Fig.~\ref{fig:causalsfho}. The mass of the red cross and the radius of the red dot in Fig.~\ref{fig:causalsfho} constitute an individual point in this diagram. The solid line provides a safe upper bound, which is employed for NS radius constraints.}
\label{fig:mmaxr16}
\end{figure}

\subsection{Constraints from causality}
It is well known that causality places constraints on stellar properties of NSs by requiring that the speed of sound $c_s$ of an EoS cannot exceed the speed of light $c$. This limits the stiffness of the EoS, and leads to $M_\mathrm{max}<\frac{1}{2.82}\frac{c^2}{G}R_\mathrm{max}$, see e.g.~\cite{Lattimer2016} for an extended discussion. Similarly, one can also find an upper limit for  $M_\mathrm{max}$ for a given radius $R_{1.6}$. We consider a sample of 23 EoSs and artificially modify those EoSs in the high-density regime. Above the central density of a 1.6~$M_\odot$ NS we replace the original EoS by the maximally stiff EoS, i.e. an EoS with $c_s=\sqrt{\frac{dp}{de}}=c$. Solving the stellar structure equations for these modified EoSs one finds the highest maximum mass $M_\mathrm{max}^\mathrm{up}$ which is compatible with a given $R_{1.6}$.

An example is given in Fig.~\ref{fig:causalsfho}. The solid black curve shows the mass-radius relation for the original SFHO EoS~\cite{Steiner2013}. The red dashed line displays the stellar configurations for the modified EoS with the maximally possible stiffening beyond the central density of a 1.6~$M_\odot$ NS. In this particular example $M_\mathrm{max}$ is increased by about 0.2~$M_\odot$ if the EoS was maximally stiff beyond the central density of a 1.6~$M_\odot$ NS.

Repeating this procedure of modifying the high-density regime for a large number of possible EoSs yields Fig.~\ref{fig:mmaxr16}. The figure shows $M_\mathrm{max}^\mathrm{up}$, which is the maximum mass of the artificially stiffened EoS, as function of $R_{1.6}$. For the SFHO EoS $M_\mathrm{max}^\mathrm{up}$ corresponds to the red cross in Fig.~\ref{fig:causalsfho}, while $R_{1.6}$ is given by the radius of the red point. The solid line is given by 
\begin{equation}\label{eq:causal16}
M_\mathrm{max}^\mathrm{up}=\frac{1}{3.1}\frac{c^2}{G}R_{1.6}
\end{equation}
and provides a robust upper limit on $M_\mathrm{max}$ for a given $R_{1.6}$. We expect only minor changes if for the same $R_{1.6}$ the EoS was modified at lower densities.

\subsection{Radius constraints}
We can now combine the arguments of (a), (b) and (c) to derive lower limits for $R_\mathrm{max}$ and $R_{1.6}$. We find
\begin{eqnarray}
M_\mathrm{tot}^\mathrm{measured}&<&\left(-3.38\frac{G\,M_\mathrm{max}}{c^2\,R_\mathrm{max}} +2.43 \right)\,M_\mathrm{max}\\
&<&\left( -\frac{3.38}{2.82} +2.34\right) \frac{1}{2.82}\frac{c^2}{G}R_\mathrm{max}.
\end{eqnarray}
In the first line we assume that the measured total binary mass in GW170817 $M_\mathrm{tot}^\mathrm{GW170817}$ is smaller than the threshold mass for prompt black hole formation $M_\mathrm{thres}$, which is given by the empirical relation Eq.~(\ref{eq:mthrrmax}). Then we insert the causality constraint to eliminate $M_
\mathrm{max}$ and find
\begin{equation}\label{eq:radmax}
R_\mathrm{max}> 2.29\,\frac{G}{c^2} \,M_\mathrm{tot}^\mathrm{measured}.
\end{equation}
Similarly, by using Eq.~(\ref{eq:mthrr16}) and Eq.~(\ref{eq:causal16}), one obtains
\begin{equation}\label{eq:rad16}
R_{1.6}> 2.55\,\frac{G}{c^2} \,M_\mathrm{tot}^\mathrm{measured}.
\end{equation}
It is very conservative to employ the limits imposed by causality because the true EoS is unlikely to be that stiff. Also, one may assume that the merger remnant survived for at least 10~ms before collapsing to a black hole. This effectively means that the binary mass of GW170817 is at least $\Delta M\approx 0.1~M_\odot$ smaller than $M_\mathrm{thres}$, which can be included in Eqs.~(\ref{eq:radmax}) and~(\ref{eq:rad16}). This finally yields $R_\mathrm{max}> 2.29\,\frac{G}{c^2} \,(M_\mathrm{tot}^\mathrm{measured}+\Delta M)=9.6$~km and $R_{1.6}> 2.55\,\frac{G}{c^2} \,(M_\mathrm{tot}^\mathrm{measured}+\Delta M)=10.7$~km. A detailed discussion of errors can be found in~\cite{Bauswein2017}. See also~\cite{Koeppel2019} for radius constraints resulting from a very similar line of arguments. Figure~\ref{fig:RM} compares our constraints with a large set of EoSs available in the literature. Based on these arguments some of the very soft EoSs can be ruled out, which complements the upper limits on NS radii arising from the measurement of finite-size effects during the late inspiral phase.

\begin{figure}\label{fig:RM}
\includegraphics[width=0.8\columnwidth]{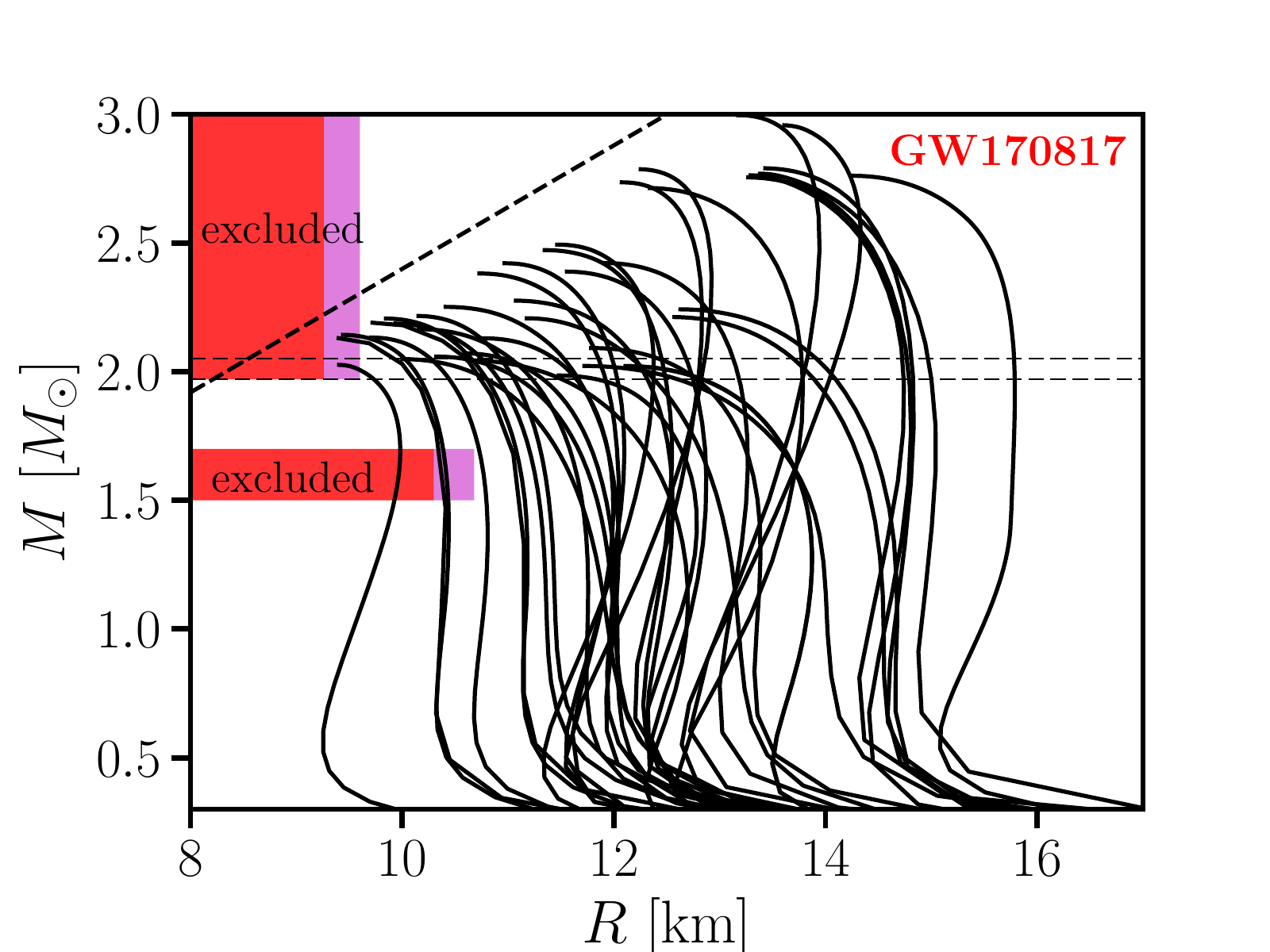}
\caption{Mass-radius relations of different EoS models available in the literature. The red areas provide a very conservative lower bound on NS radii. The cyan area display a more realistic constraint, which exclude some very soft EoSs~\cite{Bauswein2017}. The thin horizontal lines indicate most massive accurately measured NS mass~\cite{Antoniadis2013}. The thick dashed curve marks the causality limit. Figure taken from~\cite{Bauswein2017}.}
\label{fig:RM}
\end{figure}

\subsection{Discussion}
We stress some more aspects. First, any new future measurement can be employed to strengthen these radius constraints if evidence for no direct black-hole formation is found. For higher binary masses the lower limits on NS radii increase as can be easily seen from Eqs.~(\ref{eq:radmax}) and~(\ref{eq:rad16}), which can be directly applied to any new measurement. Second, if evidence for a prompt collapse of the merger remnant is found in some future event, the measured binary mass exceeds $M_\mathrm{thres}$. Following a similar line of arguments, one can then derive an upper limit on NS radii and $M_\mathrm{max}$ (see~\cite{Bauswein2017} for details). Third, these limits on NS radii rely only on the mass measurement and the interpretation of the electromagnetic signal but not on the extraction of finite-size effects of the GW signal. This implies that also future events with low signal-to-noise ratio can be employed to improve these constraints if the identification of an electromagnetic counterpart allows to infer the merger outcome. We anticipate that as more events are observed, our understanding of which of those mergers led to a prompt collapse and which did not, will increase and allow a robust distinction. Fourth, the errors of radius limits depend only on the errors of the mass measurement and the accuracy of the empirical relations for $M_\mathrm{thres}$, which can be easily propagated through the derivation (see~\cite{Bauswein2017}). Fifth, we emphasize that the use of the causality argument is very conservative since the speed of sound of the true EoS will likely be smaller than the speed of light. Also, the empirical relations Eqs.~(\ref{eq:mthrrmax}) and~(\ref{eq:mthrr16}) are derived for equal-mass mergers. Asymmetric mergers lead to threshold masses which are equal or slightly smaller than that of equal-mass systems. Hence, adopting the empirical relations for symmetric mergers is conservative as well.

For a comparison to the constraints obtained from finite-size effects during the inspiral, it is helpful to convert the above radius constraints to limits on the tidal deformability. The tidal deformability is defined by $\Lambda=\frac{2}{3}k_2\left(\frac{c^2\,R}{G\,M}\right)^5$ with the tidal Love number $k_2$, the NS mass $M$ and the NS radius $R$ (see~\cite{Hinderer2008}).  The combined tidal deformability $\tilde{\Lambda}=\frac{16}{13}\frac{(M_1+12 M_2) M_1^4\Lambda_1  + (M_2+12 M_1) M_2^4\Lambda_2 }{(M_1+M_2)^5}$ with the masses $M_{1/2}$  and tidal deformabilities $\Lambda_{1/2}$ of the individual stars is the quantity which is directly inferred from the GW signal~\cite{Abbott2017}. For equal-mass binaries the combined tidal deformability equals the tidal deformability of the individual stars. For not too asymmetric systems $\tilde{\Lambda}$ is very close to the combined tidal deformability of the equal-mass system with the same chirp mass~$\mathcal{M}=(M_1 M_2)^{3/5}(M_1+M_2)^{-1/5}$. $\tilde{\Lambda}$ was found to be smaller than $\sim 800$ with the precise value depending on priors and other assumptions~\cite{Abbott2017,TheLIGOScientificCollaboration2018,De2018}.

\begin{figure}
\includegraphics[width=0.8\columnwidth]{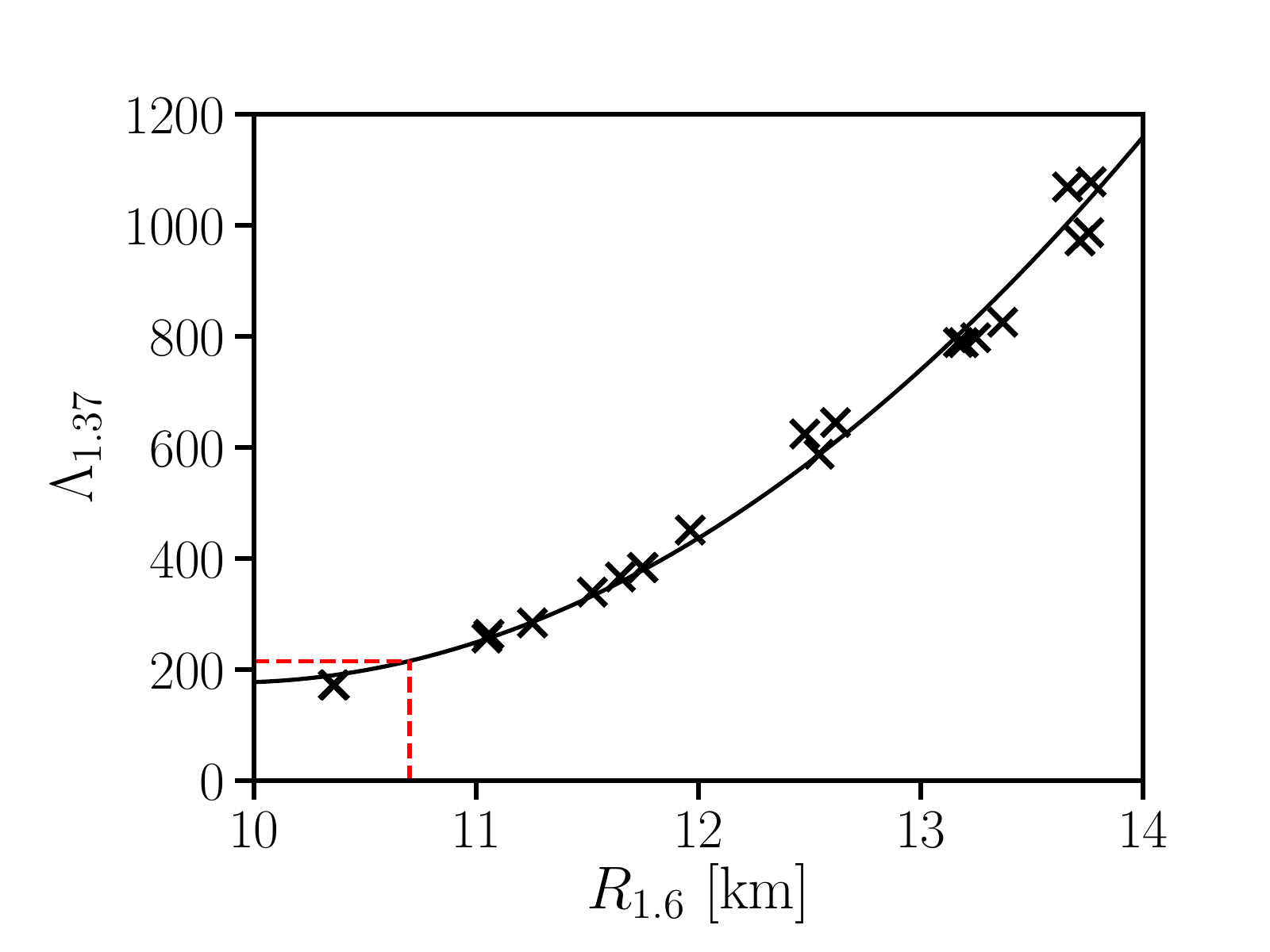}
\caption{Tidal deformability $\Lambda_{1.37}$ of a 1.37~$M_\odot$ NS as function of the radius $R_{1.6}$ of nonrotating NSs with a mass of 1.6~$M_\odot$ (crosses). The solid curve is a quadratic least-squares fit to the data. The red dashed curve indicates how the radius constraint of $R_{1.6}<10.7$~km as is discussed in the main text is converted to a limit on the tidal deformability.}
\label{fig:lam}
\end{figure}

Figure~\ref{fig:lam} displays the tidal deformability of a 1.37~$M_\odot$ NS as function of $R_{1.6}$ for a large set of candidate EoSs available in the literature. In this plot we relate stellar parameters of nonrotating stars of different masses. The relatively tight relation allows an approximate conversion of the constraint on $R_{1.6}$ to a limit on the tidal deformability $\Lambda_{1.37}$. Note that for an accurate conversion a full coverage of all possible EoSs would be required, which is beyond the scope of this work. The constraint $R_{1.6}>10.7$~km derived above corresponds to $\Lambda_{1.37}>210$ as indicated by the dashed curve in Fig.~\ref{fig:lam}.

The lower limit of $\Lambda_{1.37}\gtrsim210$ appears weaker than $\Lambda_{1.37}\gtrsim400$ from~\cite{Radice2018} (revised to $\Lambda_{1.37}\gtrsim300$ in~\cite{Radice2019}) who followed a similar argumentation and observation namely that early black-hole formation would result in a dimmer electromagnetic counterpart than that of GW170817. The different limits on $\Lambda$ are explained by the fact that Refs.~\cite{Radice2018,Radice2019} considers only 4 different EoS models to determine the boundary between dim and bright electromagnetic counterparts. Hence, in~\cite{Radice2018,Radice2019} the limit on $\Lambda_{1.37}$ is only coarsely resolved and leads in fact to an overestimation of the lower bound on $\Lambda_{1.37}$ (see e.g. the simulation for a 1.37-1.37~$M_\odot$ merger with the Sly4 EoS with $\Lambda_{1.37}\approx 338$ in~\cite{Dietrich2017a} leading to significant mass ejection and thus a presumably bright electromagnetic counterpart compatible with GW170817, or , similarly, a bright counterpart may be expected for a 1.4-1.4~$M_\odot$ merger with the APR4 EoS with $\Lambda_{1.37}\approx 281$~\cite{Hotokezaka2013} having a total mass comparable to that of GW170817). See also the discussion in~\cite{Kiuchi2019}. We thus conclude that the current data provide a lower limit of only $\Lambda\gtrsim 210$, which is found if the full EoS dependence is considered. 

We also refer to the studies in~\cite{Wang2018a} and~\cite{Coughlin2018}, which find lower limits of $\Lambda_{1.37}\gtrsim309$ and $\Lambda_{1.37}\gtrsim279$, respecively. The comparison to these limits is not straightforward as those works involve a more sophisticated interpretation of the multi-messenger observation of GW170817.

\section{Strategy for future multi-messenger observations}

Observing time in particular at large telescopes is a limited resource. It is therefore important to define an observing strategy for future searches of electromagnetic counterparts after a trigger by a GW event indicates the collision of a NS merger. This strategy should specify for which events follow-up observations promise the highest scientific gain since possibly not for all future triggers an extensive search of the electromagnetic counterpart can be initiated.

We thus discuss which potential multi-messenger events are the most promising for constraints as the one presented above. This means we focus on the determination of threshold binary mass $M_\mathrm{thres}$ for prompt black-hole formation because measuring $M_\mathrm{thres}$ provides crucial information about the properties of high-density matter. It can be employed to constrain NS radii (see Sect.~\ref{sec:main}) and to determine the unknown maximum mass $M_\mathrm{max}$ of NSs\footnote{If the radius of a 1.6~$M_\odot$ NS is known to some accuracy, the empirical relation for $M_\mathrm{thres}(M_\mathrm{max};R_{1.6})$ expressed by Eq.~(\ref{eq:mthrr16}) can be inverted to yield the maximum mass $M_\mathrm{max}$ of nonrotating NS~\cite{Bauswein2013}.} if radii are known with some precision~\cite{Bauswein2013}. Moreover, the collapse behavior, i.e. the distinction between direct and delayed collapse of the remnant, may be crucial for the development of a relativistic outflow~\cite{Ruiz2017}. Hence, knowledge of $M_\mathrm{thres}$ may be important to interpret future simultaneous observations of short gamma-ray bursts and GWs. If  $M_\mathrm{thres}$ is known, the binary mass measured through the GW inspiral phase reveals whether or not the merger resulted in a direct gravitational collapse.

We therefore propose here an observing strategy to determine $M_\mathrm{thres}$ through multi-messenger observations with a minimum number of follow-up searches. We assume that generally mergers with $M_\mathrm{tot}>M_\mathrm{thres}$ can be distinguished from systems with $M_\mathrm{tot}>M_\mathrm{thres}$ by their electromagnetic emission in the optical/infrared. Simulations suggest that direct black-hole formation leads to dim electromagnetic counterparts, while no collapse or a delayed collapse result in brighter optical emission. 

Considering these results, there is good evidence that the total mass of GW170817 is smaller than $M_\mathrm{thres}$, hence it provides a lower limit on $M_\mathrm{thres}$ (see Sect.~\ref{sec:main}). We thus argue that an event with a total binary mass larger than that of GW170817 may be more rewarding in comparison to a trigger with a lower total binary mass. A high-mass merger may provide an upper limit on $M_\mathrm{thres}$ (in case of a relatively dim electromagnetic counterpart) or strengthen the current lower limit on $M_\mathrm{thres}$ (in case of a relatively bright electromagnetic counterpart) and thus imply a stronger bound on NS radii from below.

Note that a prompt collapse event with a potentially dimmer optical emission may be particularly challenging to observe. Thus, a trigger indicating a binary mass potentially above $M_\mathrm{thres}$ may motivate an intense search with dedicated instruments and strategies to uncover particularly dim events. A merger resulting in a delayed/no collapse, i.e. with $M_\mathrm{tot}<2.74~M_\odot$, may be more easy to detect.

We also emphasize that the empirical relations for $M_\mathrm{thres}$ (Eq.~(\ref{eq:mthrr16})) and the upper limit on $M_\mathrm{max}$ for a given $R_{1.6}$ (Eq.~(\ref{eq:causal16})) in combination with the current upper limit on NS radii suggest that mergers with $M_\mathrm{tot}$ exceeding 3.73~$M_\odot$ are in any case expected to result in a prompt collapse\footnote{In detail, the current upper limit on the threshold binary mass for prompt collapse is derived as follows. $M_\mathrm{thres}$ increases with $R_{1.6}$ and with $M_\mathrm{max}$ (in the relevant regime). For a fixed $M_\mathrm{max}$ the upper bound on $R_{1.6}\lesssim R_{1.6}^\mathrm{up}=14$~km implies an upper limit on $M_\mathrm{thres}$  (Eq.~(\ref{eq:mthrr16})). Furthermore, $R_{1.6}\lesssim 14$~km yields an upper limit  $M_\mathrm{max}^\mathrm{up}(R_{1.6})$ on $M_\mathrm{max}$ through the causality constraint (Eq.~(\ref{eq:causal16}), see Fig.~\ref{fig:mmaxr16}). Combining these arguments yields an upper limit on the threshold mass of
\begin{equation}
M_\mathrm{thres}<\left(-3.606\frac{G\,M^\mathrm{up}_\mathrm{max}}{c^2\,R^\mathrm{up}_{1.6}} +2.38 \right)\,M^\mathrm{up}_\mathrm{max} =\left(\frac{-3.606}{3.1} +2.38 \right)\frac{c^2 \, R_{1.6}^\mathrm{up}}{3.1\,G}=3.73~M_\odot,
\end{equation}
where we first use Eq.~(\ref{eq:causal16}) and then insert $R_{1.6}^\mathrm{up}=14$~km, which is the limit provided by the measurement of the tidal deformability in GW170817.
}. Hence, a trigger with a binary mass above 3.73~$M_\odot$ may point to a less promising target. On the one hand the outcome may be anticipated and does not provide new constraints on the nuclear EoS, on the other hand the detection may be more challenging since the electromagnetic emission may be relatively dim. We thus suggest to first focus resources on events with a binary mass between 2.74~$M_\odot$ and 3.73~$M_\odot$. The upper limit may be further reduced by future measurements providing stronger upper bounds on NS radii.

After an upper limit on $M_\mathrm{thres}$ has been established by a future observation, the most rewarding systems will be those with a binary mass (or chirp mass) in between the lower and upper limits on $M_\mathrm{thres}$. Following this strategy, more precise estimates of the threshold binary mass will be obtained. Comparing the different future electromagnetic observations will also clarify whether or not the collapse behavior has indeed a strong impact on the properties of the optical emission as suggested by simulations.

\section{Perspective for Nuclear (Astro)physics}

The relatively early detection of a NS merger~\footnote{Early detection is meant in the sense that the first measurement succeeded before the current GW instruments reached their design sensitivity.} indicates that many more events will be observed as the sensitivity of the current generation of GW detectors increases. Future measurements similar to that of GW170817 can be employed to strengthen the constraints on NS radii and the maximum mass of NS following the procedure described above if the observation of an electromagnetic counterpart allows the inference of the immediate merger outcome. See~\cite{Bauswein2017} for a discussion of hypothetical future cases.
\begin{figure}
\includegraphics[width=0.8\columnwidth]{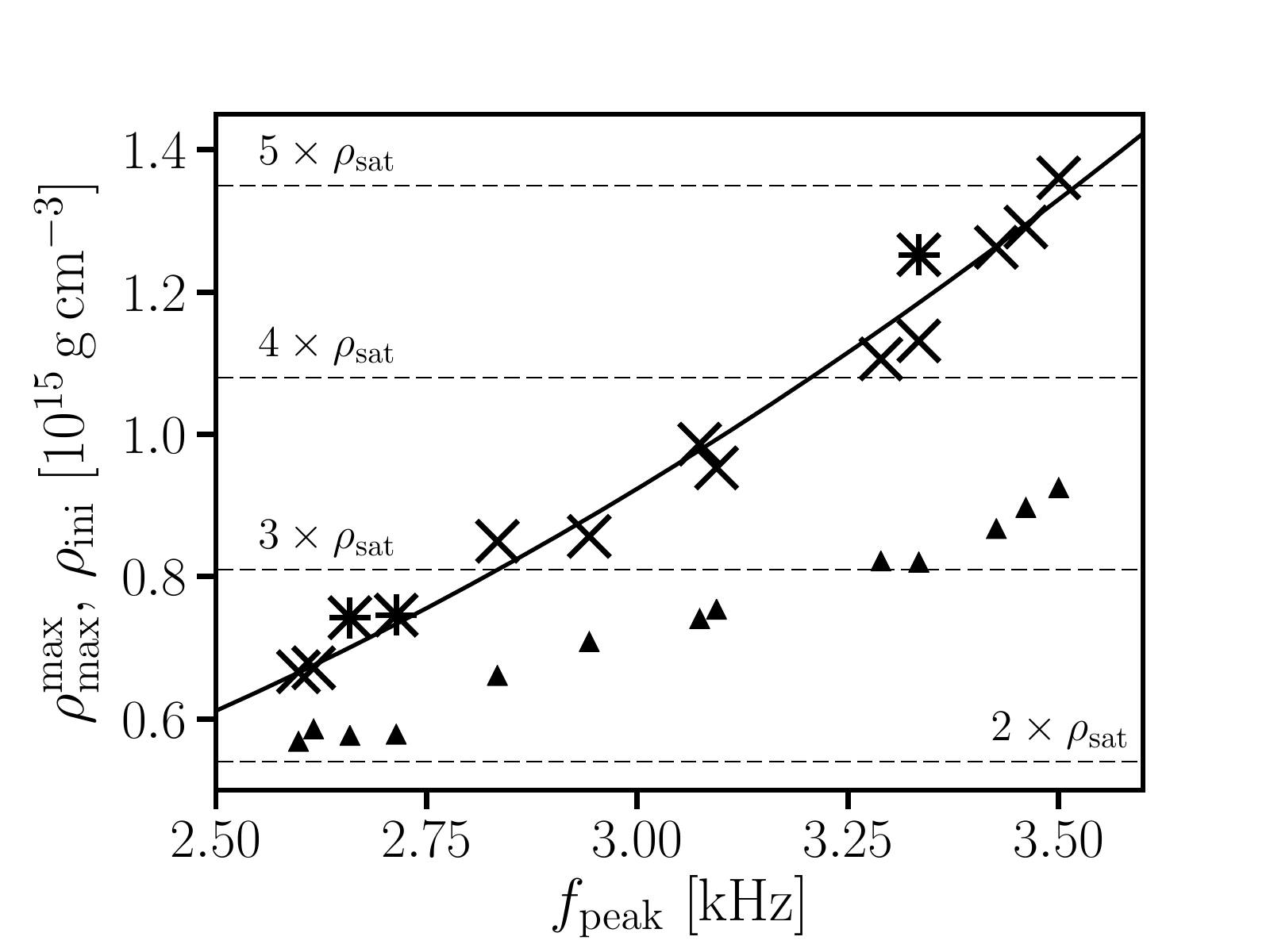}
\caption{Different rest-mass density regimes probed during the inspiral and the postmerger phase as function of the dominant postmerger GW frequency $f_\mathrm{peak}$ for 1.35-1.35~$M_\odot$ mergers for different EoS models. The triangles display the maximum density during the inspiral. The crosses and asterisks show the highest density which is reached during the early postmerger evolution. Figure adopted from~\cite{Bauswein2019}.}
\label{fig:fpeakrho}
\end{figure}

\begin{figure}
\includegraphics[width=0.8\columnwidth]{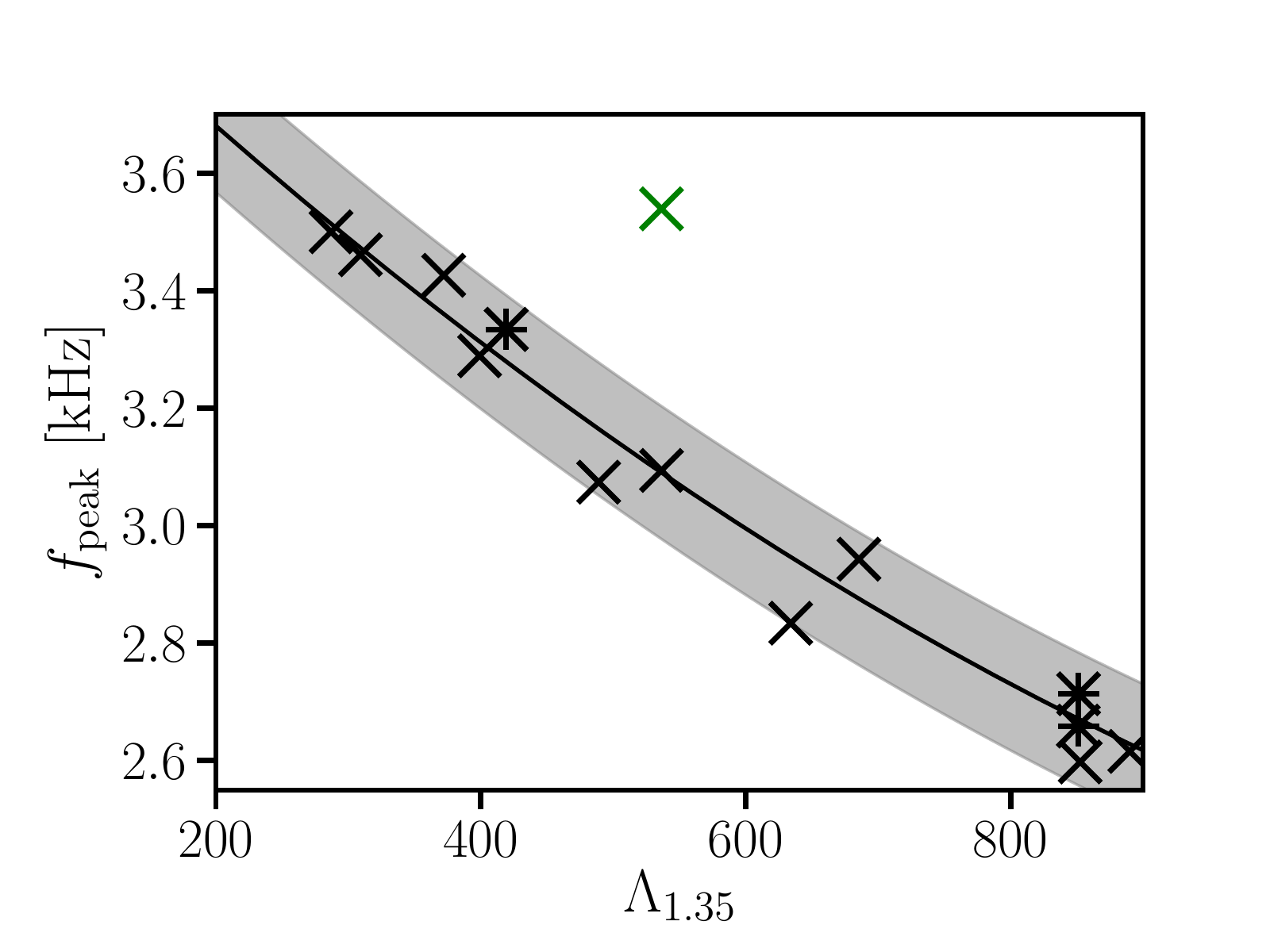}
\caption{Dominant postmerger GW frequency $f_\mathrm{peak}$ as function of the tidal deformability $\Lambda_{1.35}$ in 1.35-1.35~$M_\odot$ mergers for different EoS models. The black symbols refer to purely baryonic EoSs. The resulting correlation between $f_\mathrm{peak}$ and $\Lambda_{1.35}$ is well described by a least-squares fit (black curve) with relatively small deviations between the fit and the underlying data (black symbols). The maximum deviation is indicated by the  gray band. An EoS with a strong first-order phase transition to deconfined quark matter (green symbol) occurs as a clear outlier. Such a feature would thus provide strong evidence for the occurrence of the hadron-quark phase transition in NSs. Hyperonic EoS models are displayed by asterisks. Figure taken from~\cite{Bauswein2019}.}
\label{fig:fpeaklam}
\end{figure}

The increase of the detector sensitivity during the next observing runs also implies that the signal-to-noise ratio of an event at a distance similar to that of GW170817 will be higher. This will significantly improved constraints on the tidal deformability from the inspiral phase and will thus further constrain the EoS in the density regime which is probed by the inspiraling stars.

In addition, as the instruments approach design sensitivity the detection of GWs from the postmerger phase comes into reach for cases which do not result in a prompt gravitational collapse of the remnant. Measuring gravitational radiation from the postmerger remnant is complementary to the inference of finite-size effects during the inspiral phase in a twofold sense. First, measuring the dominant oscillation frequency of the postmerger NS remnant will provide an independent measurement on NS radii~\cite{Bauswein2012,Bauswein2012a,Hotokezaka2013a,Takami2014,Bernuzzi2015,Bauswein2015,Bauswein2016}. The complementarity concerns the underlying physical model for the interpretation of the data and the employed data analysis methods, e.g.~\cite{Chatziioannou2017}. Second, the postmerger remnant involves higher densities than the progenitor stars. This can be seen in Fig.~\ref{fig:fpeakrho}, where the triangles refer to the highest density during the inspiral phase, while the crosses and asterisks mark the maximum density of the postmerger stage for 1.35-1.35~$M_\odot$ mergers. Simulations with different EoSs yield different postmerger GW frequencies $f_\mathrm{peak}$, which characterize the stiffness of the EoSs. Soft EoSs yield high frequencies and lead to a particularly strong increase of the density in the postmerger phase. It is obvious that GWs from the postmerger stage probe a different regime of the nuclear EoSs. In particular, the detection of postmerger GW emission offers the possibility to learn about the presence of a hadron-quark phase transition in compact stars~\cite{Bauswein2019,Most2019}. Figure~\ref{fig:fpeaklam} reveals that the presence of a strong first-order phase transition results in a significant increase of the dominant postmerger GW frequency $f_\mathrm{peak}$ for a given tidal deformability, which is measured from the inspiral phase. See e.g.~\cite{Paschalidis2018,Drago2018,Han2019} for other studies of quark stars in the context of binary mergers.

Finally, we emphasize that observational evidence for the presence or absence of postmerger GW emission in combination with the binary mass measurements from the inspiral phase can be employed to directly determine the threshold binary mass for prompt black-hole formation. As discussed above (see Eqs.~(\ref{eq:mthrrmax}) and~(\ref{eq:mthrr16})) this information can be used to obtain the maximum mass of nonrotating NSs~\cite{Bauswein2013}, which would thus provide very important constraints on the very high density regime of the nuclear EoS.

{\it Acknowledgments: The author acknowledges support by the European Research Council (ERC) under the European Union's Horizon 2020 research and innovation programme under grant agreement No. 759253, the Klaus-Tschira Foundation, and the Sonderforschungsbereich SFB 881 ``The Milky WaySystem'' (subproject A10) of the German Research Foundation (DFG). I thank the organizers of the workshop ``Nuclear astrophysics in the new era of multi-messenger astronomy''.}

\section*{References}


\end{document}